# Spatially resolved electronic structure of twisted graphene

Qirong Yao[1]*, Rik van Bremen[1], Guus J. Slotman[2], Lijie Zhang[1], Sebastiaan Haartsen[1], Kai Sotthewes[1], Pantelis Bampoulis[1], Paul L. de Boeij[1], Arie van Houselt[1], Shengjun Yuan[3,2], and Harold J.W. Zandvliet[1]*

[1]Physics of Interfaces and Nanomaterials and MESA+ Institute for Nanotechnology, University of Twente, P.O. Box 217, 7500AE Enschede, The Netherlands

[2]Institute for Molecules and Materials, Radboud University, Heijendaalseweg 135, 6525AJ Nijmegen, The Netherlands

[3]School of Physics and Technology, Wuhan University, Wuhan, 430072, China

We have used scanning tunneling microscopy and spectroscopy to resolve the spatial variation of the density of states of twisted graphene layers on top of a highly oriented pyrolytic graphite substrate. Owing to the twist a moiré pattern develops with a periodicity that is substantially larger than the periodicity of a single layer graphene. The twisted graphene layer has electronic properties that are distinctly different from that of a single layer graphene due to the nonzero interlayer coupling. For small twist angles (~$1°$-$3.5°$) the integrated differential conductivity spectrum exhibits two well-defined Van Hove singularities. Spatial maps of the differential conductivity that are recorded at energies near the Fermi level exhibit a honeycomb structure that is comprised of two *inequivalent* hexagonal sub-lattices. For energies $|E-E_F|>0.3$ eV the hexagonal structure in the differential conductivity maps vanishes. We have performed tight-binding calculations of the twisted graphene system using the propagation method, in which a third graphene layer is added to mimic the substrate. This third layer lowers the symmetry and explains the development of the two hexagonal sub-lattices in the moiré pattern. Our experimental results are in excellent agreement with the tight-binding calculations.

*) Corresponding authors: q.yao@utwente.nl and h.j.w.zandvliet@utwente.nl

**Introduction**

The discovery of graphene in 2004 by Novoselov and Geim [1,2] has resulted into a long list of exciting and beautiful discoveries, as well as a new research field that deals with two-dimensional materials [3-10]. Graphene is a semimetal, i.e. the material is gapless, but the density of states vanishes at the Fermi level [3]. The electronic band structure of a single graphene layer near the Fermi level is characterized by linearly dispersing energy bands that form Dirac cones at the K and K′ points of the Brillouin zone [3]. The apex of these cones is located at the Fermi level. When two layers of graphene are stacked on top of each other the electronic structure alters substantially. The low energy electronic band structure of bilayer graphene depends on how the two graphene layers are stacked [5]. The most common stacking is the so-called AB or Bernal stacking. The atoms of one of the hexagonal sub-lattices of the top layer ($A_1$) are located on-top of the atoms of one of the sub-lattices of the bottom layer ($B_2$). The other atoms ($B_1$ and $A_2$) do not lie directly below or above an atom of the other layer. Highly oriented pyrolytic graphite is often stacked in the Bernal configuration. A small twist angle of the top graphene layer with respect to the second graphene layer results into a so-called moiré pattern. The periodicity of this moiré pattern depends on the exact value of the twist angle. The electronic structure of this moiré pattern is characterized by a set of two Dirac cones that are located close to each other in reciprocal space. The crossing of these two Dirac cones results into two Van Hove singularities.

In 2010 Li et al. [11] used scanning tunneling microscopy and spectroscopy to analyze these Van Hove singularities in twisted graphene layers. For small twist angles these authors observed two well-defined Van Hove singularities, one located just above the Fermi level and the other one located just below the Fermi level. The experimentally determined energy separation between these two Van Hove singularities nicely agrees with tight-binding calculations, provided that reasonable assumptions for the hopping integrals are made [11]. In addition, the authors pointed out that the two Van Hove singularities can become asymmetric (in position with respect to the Fermi level and amplitude) due to the presence of an interlayer bias. This interlayer bias is caused by the potential that is applied across the scanning tunneling microscopy junction. In the scanning tunneling microscopy data by Yin et al. [12] a similar asymmetry and shift was found and discussed. Yan et al. [13] studied the angle-dependent van Hove singularities and found, in contrast to predictions by band structure calculations, that the Fermi velocity is very comparable to the Fermi velocity of monolayer graphene. In a follow-up study Yan et al. [14] showed the breakdown of van Hove singularities beyond a twist angle of about 3.5°, indicating that the continuum models are no longer applicable at these relatively large twist angles. Yin et al. [15] showed that there is a magic twist angle of 1.11° at which the two van Hove singularities merge together and form a well-defined peak at the charge neutrality point. In addition to this strong peak at the charge neutrality point these authors also found a set of regularly spaced peaks. These regularly spaced peaks are confined electronic states in the twisted bilayer graphene. The energy spacing of 70 meV (=$v_F$/D) agrees well with the periodicity of the moiré pattern. In another study Yin et al. [12] demonstrated that tilt grain boundaries can severely affect the structural and electronic properties of graphene multilayers. They also pointed out that tilt grain boundaries in trilayer graphene can result in the coexistence of massless Dirac fermions and massive chiral fermions. Wong et al. [16] performed local spectroscopy on gate-tunable twisted bilayer graphene. The twisted graphene bilayer was positioned on a hexagonal boron nitride substrate. Wong et al. [16] found, besides the coexistence of moiré patterns and moiré super-superlattices, also a very rich an interesting electronic structure. Despite the fact that the electronic structure of twisted bilayer graphene has been extensively studied [5,11-24], the spatial variation of electronic structure within the unit cell of the moiré pattern did not receive a lot of attention yet.

Here we have studied the spatial variation of the electronic structure of twisted graphene on highly oriented pyrolytic graphite substrates. In agreement with previous studies we found the development of two Van Hove singularities in the density of states. Spatial maps of the differential conductivity of the moiré pattern near the Fermi level reveal a honeycomb structure that is comprised of two inequivalent interpenetrating hexagonal sub-lattices. At large energies, i.e. $|E-E_F|>0.3eV$, the difference in the density of states of the two hexagonal sub-lattices fades away. Here we show that the inequivalence of these two sub-lattices can be understood if one takes into account a lowering of the symmetry due to the presence of the substrate. We will model this by introducing a third graphene layer. The fact that the spatial variation of the differential conductivity fades away at high energies hints to an electronical instability.

**Experimental**

The experiments are performed with an ultra-high vacuum scanning tunneling microscope (Omicron). The base pressure of the ultra-high vacuum system is 1x10$^{-11}$ mbar. Before insertion of the ZYA quality highly oriented pyrolytic graphite (HOPG) substrates into the load lock of the ultra-high vacuum system we have removed several graphene layers via mechanical exfoliation using the Scotch-tape method. In order to remove any residual water from the highly oriented pyrolytic graphite surfaces we have baked the load-lock system for 24 hours at a temperature of 120 °C. After cooling down, the samples are transferred to the main chamber and subsequently inserted into the scanning tunneling microscope for imaging .

The scanning tunneling microscopy images are recorded in the constant current mode. Scanning tunneling spectroscopy spectra are recorded in two ways. In the first method we record current-voltage (*IV*) curves at many locations of the surface with the feedback loop of the scanning tunneling microscope disabled. The *dI/dV* spectra are obtained by numerical differentiation of the *IV* traces. In the second method a small sinusoidal voltage with a small amplitude of a few mV and a frequency of 1.9 kHz is added to the bias voltage. A lock-in amplifier is used to record the *dI/dV* signal.

**Theoretical**

The theoretical calculations have been performed within the framework of the Slater-Koster tight-binding model, in which we took into account the intralayer and interlayer hoppings between the $p_z$ orbitals. The nearest intralayer hoppings in all layers are fixed as $t$ = 3 eV, and the interlayer hopping between two sites in different layers is given by,

$$t_\perp = \cos^2 \alpha \; V_\sigma + \sin^2 \alpha \; V_\pi, \qquad (1)$$

where the orbital overlap is modeled as function of the angle $\alpha$ between the line connecting the two sites and the normal of the graphene plane, while $V_\sigma$ and $V_\pi$ are Slater-Koster integrals depending on the distance between the two sites. Both $V_\sigma$ and $V_\pi$ decay rapidly when the distance between the two sites is larger than the lattice parameter $a_0$ = 2.46 Å, and the contribution of $V_\pi$ is negligible in the interlayer hoppings in multilayer graphene [23,24]. Here we use 0.24 eV as the maximum value of $V_\sigma$ (for two sites with A-A stacking, the same value as used in Ref. [11]), and consider the screening effects following the environment-dependent tight-binding model introduced in Eq. (1) of Ref. [24]. The values of seven parameters fitted for the screening in multilayer graphene are taken from Ref. [24] as $\alpha_1$ = 6.175, $\alpha_2$ = 0.762, $\alpha_3$ = 0.179, $\alpha_4$ =1.411, $\beta_1$ =6.811, $\beta_2$ = 0.01, $\beta_3$ = 19.176. All the neighboring pairs within a maximum in-plane distance of 2 Å are included in the Hamiltonian.

The electronic properties such as the density of states and quasi-eigenstates, which have the real-space profiles comparable to the experimental STM results, are calculated by using the tight-binding propagation method (TBPM) [25,26]. TBPM has the advantages that the physical properties are extracted directly from the time-evolution of the wave function, without any diagonalization of the Hamiltonian matrix.

**Results and discussion**

The electronic structure of twisted bilayer graphene, where the top layer is twisted by an angle $\theta$ with respect to the bottom layer, depends on the exact value of $\theta$. Commensurably twisted bilayer graphene can result in two different moiré lattice types [27]. The first type has a simple two-dimensional hexagonal superlattice, which is similar to the AB-stacked (Bernal) lattice. The other type has a two-dimensional honeycomb superlattice comprising two equivalent hexagonal superlattices, and is similar to the AA-type stacked lattice. The honeycomb cases can be generated by twisting the two layers relative to one another over special angles $\theta$ obtained from the relation [27,28],

$$\cos(\theta) = \frac{2n^2 + 2nm - m^2}{2(n^2 + nm + m^2)}, \quad (2)$$

in which the integers $n$ and $m$ have no common divisors, and $n-m$ is not an integer multiple of 3 [11]. The superlattice vectors are then given by $A_1 = na_1 + ma_2$ respectively $A_2 = -ma_1 + (n+m)a_2$ with a supercell size factor $N = n^2 + nm + m^2$ larger then in graphene [29,30]. The simple hexagonal lattice type can be obtained from the same relations by twisting over the special angles $\theta + \pi$. As an example we show in Figure 1 a scanning tunneling microscopy image of a 2.3° twisted graphene layer on a highly oriented pyrolytic graphite surface recorded at 77 K. The periodicity of the moiré pattern is 6.2 nm ($a_0/2\sin(\theta/2)$, where $a_0$ =0.246 nm is the lattice constant of graphene [29,30]).

The twist of the top graphene layer leads to a shift of the Dirac points in momentum space [5,11]. The shift in momentum space with respect to the K point of a single layer graphene, $\Delta K$, is given by,

$$\Delta K = \pm K \sin(\theta/2) \quad (3)$$

In Figure 2(c) a schematic diagram of the energy bands of twisted bilayer graphene near the K point is shown. The crossing of the Dirac cones leads to a 'flat' region in the energy dispersion relation and thus to a divergence in the density of states, also referred to as a Van Hove singularity [31]. A prerequisite for the formation of Van Hove singularities in bilayer graphene is the presence of a non-zero interlayer coupling. For a vanishing interlayer coupling the electronic band structure of bilayer graphene reduces to that of the combination of the two independent graphene layers.

In the inset of Figure 2(a) a scanning tunneling microscopy image of a twisted graphene layer is shown. The image is recorded at room temperature and the moiré pattern has a periodicity of 7.0 nm, corresponding to a twist angle of 2.0°. The differential conductivity, *dI/dV*, which is proportional to the density of states for small biases, is depicted in Figure 2(a). The *dI/dV* spectra are obtained by numerical differentiating 3600 *I(V)* curves recorded at a 60x60 grid of the surface displayed in the inset of Figure 2(a). Two well-defined peaks are found at energies of - 110 meV and 15 meV with respect to the Fermi level, respectively. These two peaks are Van Hove singularities. At the high regions of the moiré pattern the peaks have a higher intensity as compared to the lower regions of the moiré pattern. The energy separation, relative strength and asymmetry are in good agreement with Ref. [11]. The *dI/dV* spectra in Figure 2(a) are recorded at room temperature and therefore these peaks are much broader than the peaks that are reported in Ref. [11], which are taken at 4 K. Spatial maps of the *dI/dV* recorded at various energies are shown in the two middle panels of Figure 3. In order to understand the experimental observations shown in Figure 2, we have performed theoretical calculations of the density of states by using the Slater-Koster tight-binding model for rotated bilayer and trilayer graphene, i.e., a rotated graphene layer on top of a single-layer or bilayer graphene, respectively. The numerical results of the integrated density of states are plotted in Figure 2(b). It is clear that although the two Van Hove singularities are always present when there is a rotated graphene layer, one has to take into account the third layer in order to reproduce the significant electron-hole asymmetry and the finite density of states in the vicinity of the Fermi level. The electron-hole asymmetry is enhanced if the interlayer hoppings between the top and the third layer are also included. Furthermore, by turning on the direct interactions between the top and the third layer, the whole energy spectrum is shifted to the hole direction, similar to the experimental observations. Here we want to emphasize that for a heterostructure consisting of a rotated graphene layer on top of graphite, it is not sufficient to only consider a rotated bilayer graphene in the theoretical studies. The influence of the third layer, either indirectly via the hoppings to the middle layer, or directly via the interactions to the top layer, is not negligible. It is therefore necessary to consider at least three layers in the calculations of the electronic structure and physical properties.

In Figure 4(a) high resolution spatial map of the differential conductivity of the strongest Van Hove singularity, which is located at -110 meV, is shown. This spatial map is recorded with a lock-in amplifier (modulation voltage 20 mV and frequency 1.9 kHz). The spatial *dI/dV* map exhibits atomic resolution. Even the periodicity of the top graphene layer with a lattice constant of 0.246 nm is visible. For the sake of clarity we have inverted the color scale in Figure 4(a), so dark regions refer to a high *dI/dV* signal, whereas bright spots refer to a low *dI/dV* signal. The honeycomb structure consists of two interpenetrating hexagonal sub-lattices. One hexagonal sub-lattice displays a substantially higher *dI/dV* signal than the other hexagonal sub-lattice. The occurrence of these two hexagonal sub-lattices can be understood if one takes into account a third graphene layer that breaks the symmetry of a twisted bilayer graphene. The dominant stacking arrangement of HOPG is the Bernal (AB) stacking. Consequently, half of the carbon atoms of the second graphene layer are located on top of a carbon atom of the bottom layer, whereas the other half of the second layer carbon atoms do not have a carbon atom underneath them. In Figure 4(b) the schematic diagram of trilayer graphene is depicted: the two bottom graphene layers are AB stacked, whereas the top graphene layer is twisted by 2.0° with respect to the second graphene layer. The honeycomb lattice of the moiré pattern in Figure 4(b) is composed of two interpenetrating hexagonal sub-lattices. The highest *dI/dV* signal is observed if the atoms in the second layer have atoms directly underneath them.

In the two middle panels of Figure 3 spatial maps of the differential conductivity are shown at various energies. The moiré pattern is present in the differential conductivity maps that are recorded near the Fermi level, but the structure fades away at larger energies. Also this observation is consistent with Ref. [11], albeit the sample bias range where we observe the moiré pattern in the *dI/dV* signal is substantially larger. In order to understand this strong energy dependence of the differential conductivity maps we have performed tight-binding calculations of a quasi-eigenstate, a superposition state of all degenerate eigenstates at a given energy [24]. The real-space distribution of the wave amplitude in a quasi-eigenstate corresponds to the local density of states measured in the scanning tunneling microscopy experiments [25]. In the left and right panels of Figure 3, we show contour plots of several quasi-eigenstates for a layer of rotated graphene stacked on top of an AB-stacked bilayer graphene. Our theoretical calculations of this heterostructure consisting of three graphene layers show exactly the same tendency as the experimental data, i.e. the hexagonal structure in the density states is only present near the Fermi level and fades away at higher energies.

The fact that the differential conductivity only exhibits a density modulation near the Van Hove singularities is reminiscent for a charge density wave. One of the hallmarks of a charge density wave is that the electron density and the lattice positions are coupled. Charge density waves may be generated by an exchange-driven instability of a metallic Fermi surface (Fermi nesting), or by a lattice-dynamical instability leading to a static periodic lattice distortion. It is important to point out here that a periodic potential in a Dirac system will not result in the opening of a band gap, but rather in the creation of new Dirac points and Van Hove singularities [32,33]. The concept of charge density of waves needs therefore to be revisited for Dirac systems. The energy dependent electron density modulation that we measured for twisted graphene can be fully explained by tight-binding calculations. Since electron-phonon coupling is not included in these tight-binding calculations it remains to be seen whether we are dealing here with a charge density wave.

**Conclusions**

Spatially resolved scanning tunneling spectroscopy measurements of twisted graphene reveal a hitherto unnoticed variation of the density of states within the unit cell of the moiré pattern. A honeycomb pattern is found that is comprised of two *inequivalent* hexagonal sub-lattices. The symmetry of the honeycomb lattice of the moiré pattern is broken by a third graphene layer that is stacked in a Bernal configuration with respect to the second graphene layer. Our experimental findings are in excellent agreement with tight-binding calculations.

**Acknowledgements**

We wish to thank Annalisa Fasolino, Mikhail Katsnelson and Alexander Rudenko for useful discussions. LZ and QY thank the China Scholarship Council for financial support. PB thanks the Nederlandse Organisatie voor Wetenschappelijk Onderzoek (NWO, STW 11431) for financial support. RvB and HJWZ thank the stichting voor Fundamenteel Onderzoek der Materie (FOM, FV157) for financial support. GJS and SY thank the European Research Council for financial support (Advanced Grant program, contract 338957). The support by the Netherlands National Computing Facilities foundation (NCF) are acknowledged.

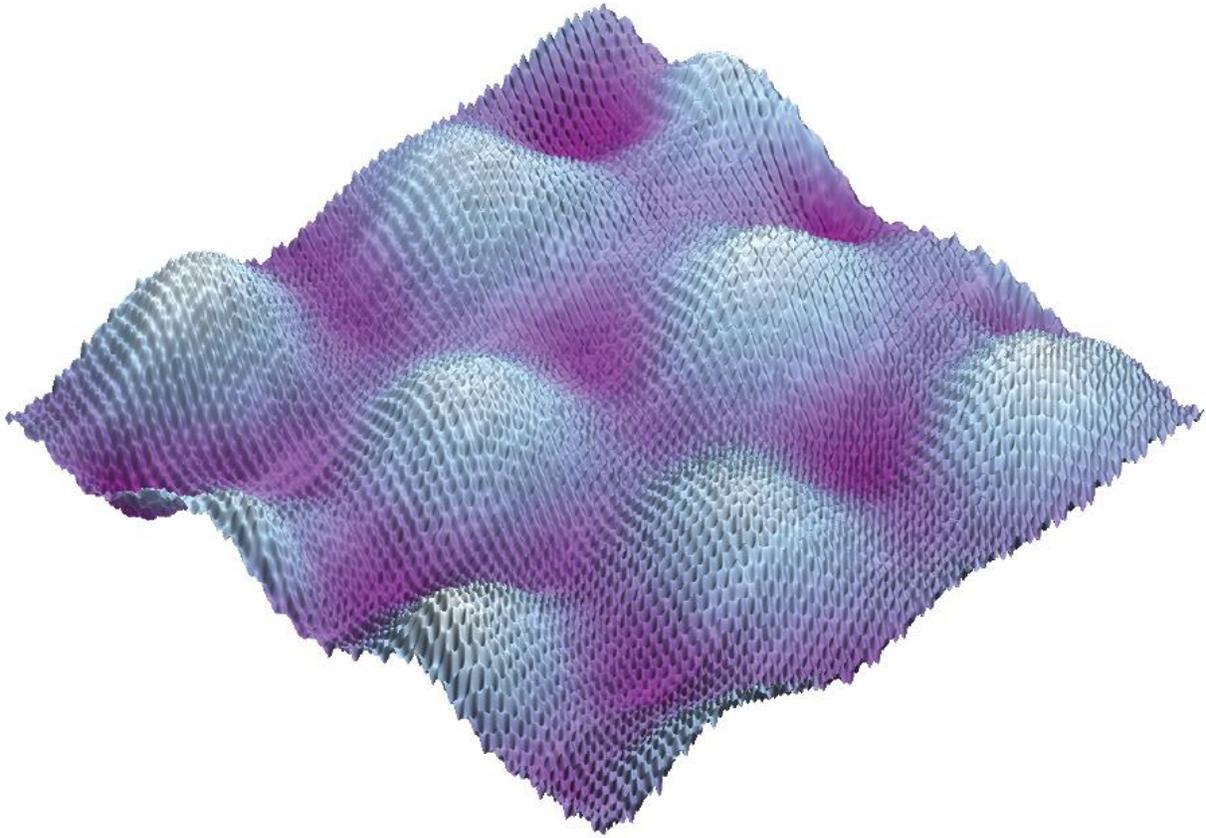

**Figure 1**
Scanning tunneling microscopy image of twisted graphene. The periodicity of the moiré pattern is 6.2 nm, corresponding to a twist angle of 2.3°. The peak-to-valley variation is 1.9 Å. Image size 15 nm x 15 nm, sample bias -1.2 V and tunnel current 400 pA. T= 77 K.

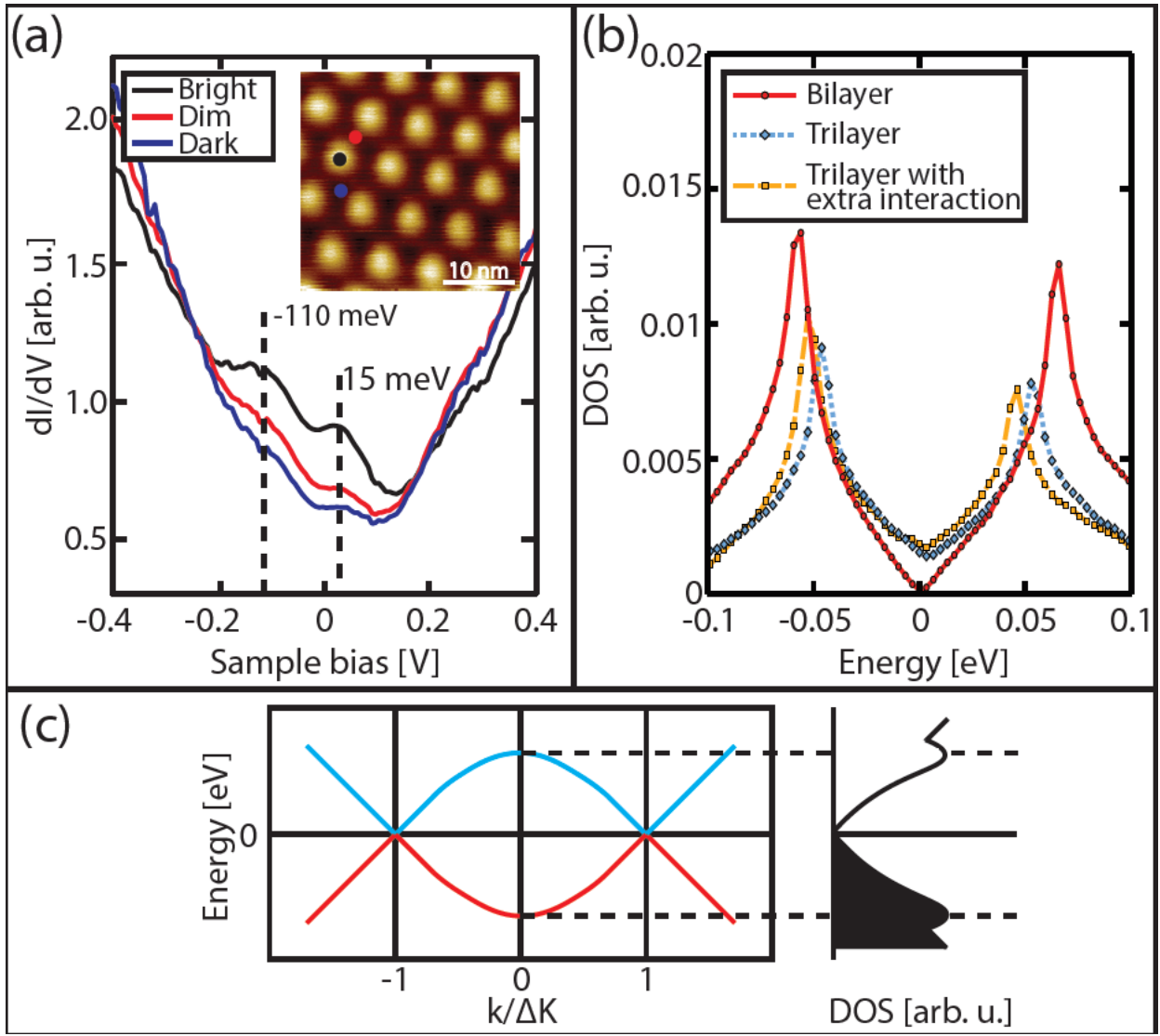

**Figure 2**
(a) Differential conductivity recorded at different locations of the STM image shown in the inset. (b) Calculated total density of states for rotated bilayer and trilayer graphene ($\theta=2.0°$, the supercell is constructed as in Ref. 9 (Eq. 1) using *m*=1 and *n*=49). For trilayer graphene with an extra interaction interlayer hoppings between the top and bottom layers with a maximum value of 0.1 eV are included (for two sites on A-A stacking). (c) Cartoon of the electronic structure of twisted bilayer graphene near the Fermi level. Red curve: energy dispersion below the Fermi level. Blue curve: energy dispersion above the Fermi level.

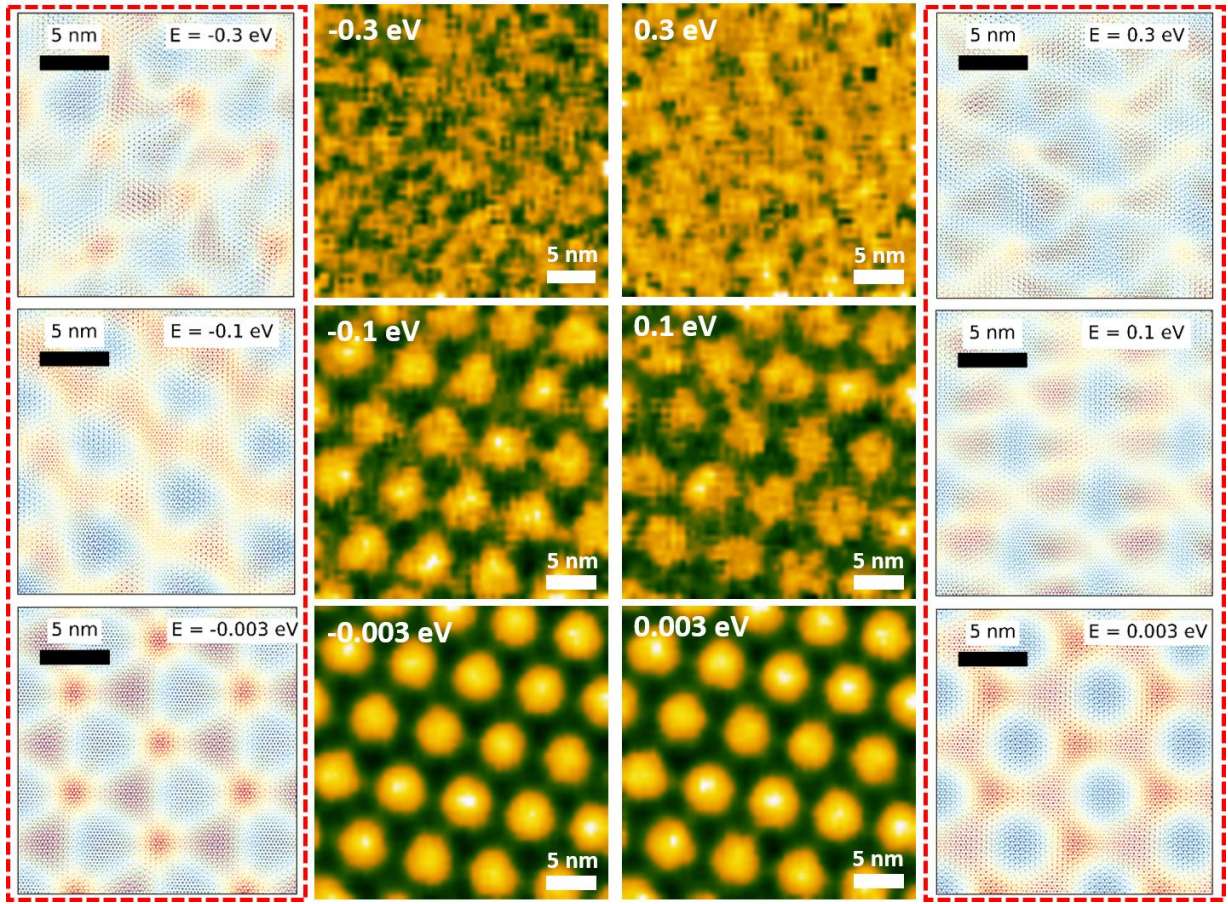

**Figure 3**

Middle panels: Spatial map of the differential conductivity at different bias voltages. The large bright spots in the *dI/dV* maps correspond to the higher parts of the moiré pattern (see Figure 1(b)). Left and right panels (in red dashed box): The real-space amplitude (logarithmic scale) of the calculated quasi-eigenstates for trilayer graphene with twisted top layer (θ=2.0°). The results are obtained by averaging over 24 initial states to mimic the randomness introduced by the initial state. In each figure, blue and red correspond to the maximal and minimal intensity, respectively. For higher absolute energy this amplitude is lower.

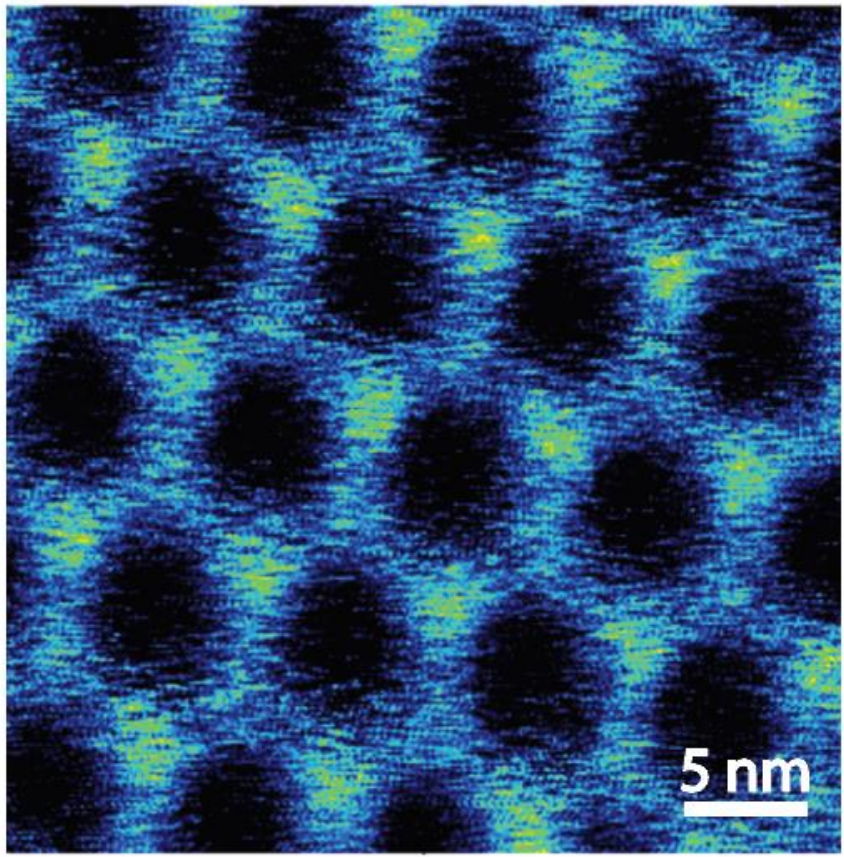

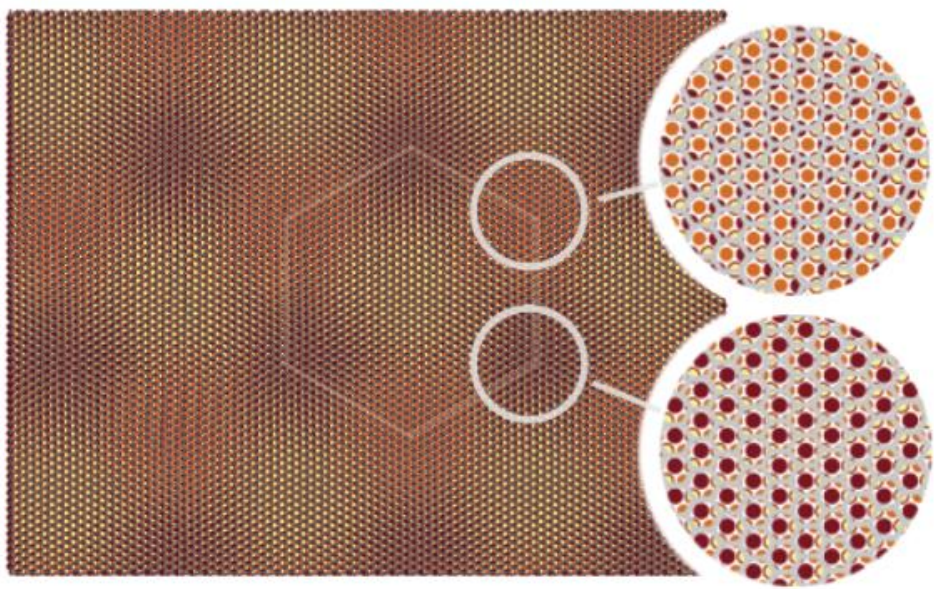

| | | | |
|---|---|---|---|
| ⬡ top layer 2° twisted | highlighting layer B on top of A (moiré pattern) | middle layer B | bottom layer A |

**Figure 4**
(a) Spatial map of the differential conductivity. For the sake of clarity we have inverted the color scale. The large dark spots correspond to the bright spots of the moiré pattern in topography image. The *dI/dV* maps are recorded with a lock-in technique (sample bias -0.3 V, modulation voltage 20 mV and frequency lock-in amplifier 1.9 kHz).
(b) Structural model of trilayer graphene. The bottom two layers are stacked in the Bernal configuration, whereas the top layer graphene is twisted by 2.0° with respect to the second graphene layer.